\documentclass[slac_one]{revtex4}
\usepackage{graphicx}
\usepackage{fancyhdr}
\begin{document}

\title{Initial Hubble Diagram Results from the Nearby Supernova Factory}

%

\newcommand{\lpnhe}{\affiliation{LPNHE, Paris, France}}
\newcommand{\cral}{\affiliation{CRAL, Lyon, France}}
\newcommand{\ipnl}{\affiliation{IPNL, Lyon, France}}
\newcommand{\lbl}{\affiliation{LBL, Berkeley, CA}}
\newcommand{\yale}{\affiliation{Yale University, New Haven, CT}}
\newcommand{\etal}{{\it et al.}}

\author{S. Bailey}\lpnhe
\author{G. Aldering}\lbl
\author{P. Antilogus}\lpnhe
\author{C. Aragon}\lbl
\author{C. Baltay}\yale
\author{S. Bongard}\lbl
\author{C. Buton}\ipnl
\author{M. Childress}\lbl
\author{Y. Copin}\ipnl
\author{E. Gangler}\ipnl
\author{S. Loken}\lbl
\author{P. Nugent}\lbl
\author{R. Pain}\lpnhe
\author{E. Pecontal}\cral
\author{R. Pereira}\lpnhe
\author{S. Perlmutter}\lbl
\author{D. Rabinowitz}\yale
\author{G. Rigaudier}\cral
\author{P. Ripoche}\lpnhe
\author{K. Runge}\lbl
\author{R. Scalzo}\yale
\author{G. Smadja}\ipnl
\author{C. Tao}\ipnl
\author{R. C. Thomas}\lbl
\author{C. Wu}\lpnhe

\begin{abstract}
The use of Type~Ia supernovae as distance indicators led to the
discovery of the accelerating expansion of the universe a decade ago.
Now that large second generation surveys have significantly increased the
size and quality of the high-redshift sample, the cosmological constraints
are limited by the currently available sample of $\sim$50 cosmologically
useful nearby supernovae.  The Nearby Supernova Factory addresses this
problem by discovering nearby supernovae and observing their
spectrophotometric time development.
Our data sample includes over 2400 spectra from spectral timeseries
of 185 supernovae.
This talk presents results from a portion of this sample
including a Hubble diagram (relative distance {\it vs.} redshift)
and a description of some analyses using this rich
dataset.\footnote{This is an abridged version of this paper,
trimmed to meet the page length requirements of the ICHEP08 proceedings;
see arXiv:0810.3499v1 for a more detailed writeup.}
\end{abstract}

\maketitle

\thispagestyle{fancy}


\section{INTRODUCTION}

The method of using supernovae to constrain dark energy
hinges upon comparing the brightnesses of distant and nearby
supernovae.
By using the ratio of luminosities of distant and nearby supernovae, the
uncertainties in the Hubble constant $H_0$ and the absolute magnitude of
Type~Ia supernovae (SNe~Ia) are cancelled, allowing measurements of
the dark energy fraction $\Omega_\Lambda$
and the equation of state parameter $w$.
Through the success of multiple observational programs,
the high redshift supernova sample is now considerably larger than the nearby
sample \cite{scpunion}.
In addition to the statistical imbalance of the low- and high-redshift
samples, the cosmology fits are now dominated by systematic errors
related to the intercalibration of the nearby and distant samples 
and the lightcurve modeling of supernovae.
Fortunately, these issues are addressable through improved measurements
of nearby supernovae.

\section{THE NEARBY SUPERNOVA FACTORY}

The Nearby Supernova Factory (SNfactory) is a program to discover, observe
and analyze nearby supernovae to improve both the statistical and systematic
limits of cosmological measurements using supernovae.  It is comprised of
two main components: a large area supernova search followed by
spectrophotometric observations of the development of the supernovae.

The search uses the 112 CCD QUEST-II
camera \cite{questII} on the Palomar Oschin 1.2-m telescope.
The observing pattern covers
350 to 850 square degrees per night, repeating fields with a median
cadence of 5 days.
Coadded stacks of previous observations are
subtracted from the new images and remaining objects are identified
and ranked as possible supernovae.  Selected candidates are screened
using the SuperNova Integral Field Spectrometer (SNIFS) \cite{snifs},
mounted on the University of Hawaii 2.2-m telescope on Mauna Kea.
Type~Ia supernovae discovered before or at
maximum brightness with redshift $0.03 < z < 0.08$ are followed with
additional measurements by SNIFS every 2-3 nights until they are 45 days past
maximum brightness (near the end of their lightcurve, observations
are obtained less frequently).  This redshift range limits the
uncertainties from peculiar velocities while maximizing the lever arm
for cosmology fits with the high redshift sample.


\begin{figure*}[t]
\centering
\includegraphics[width=65mm]{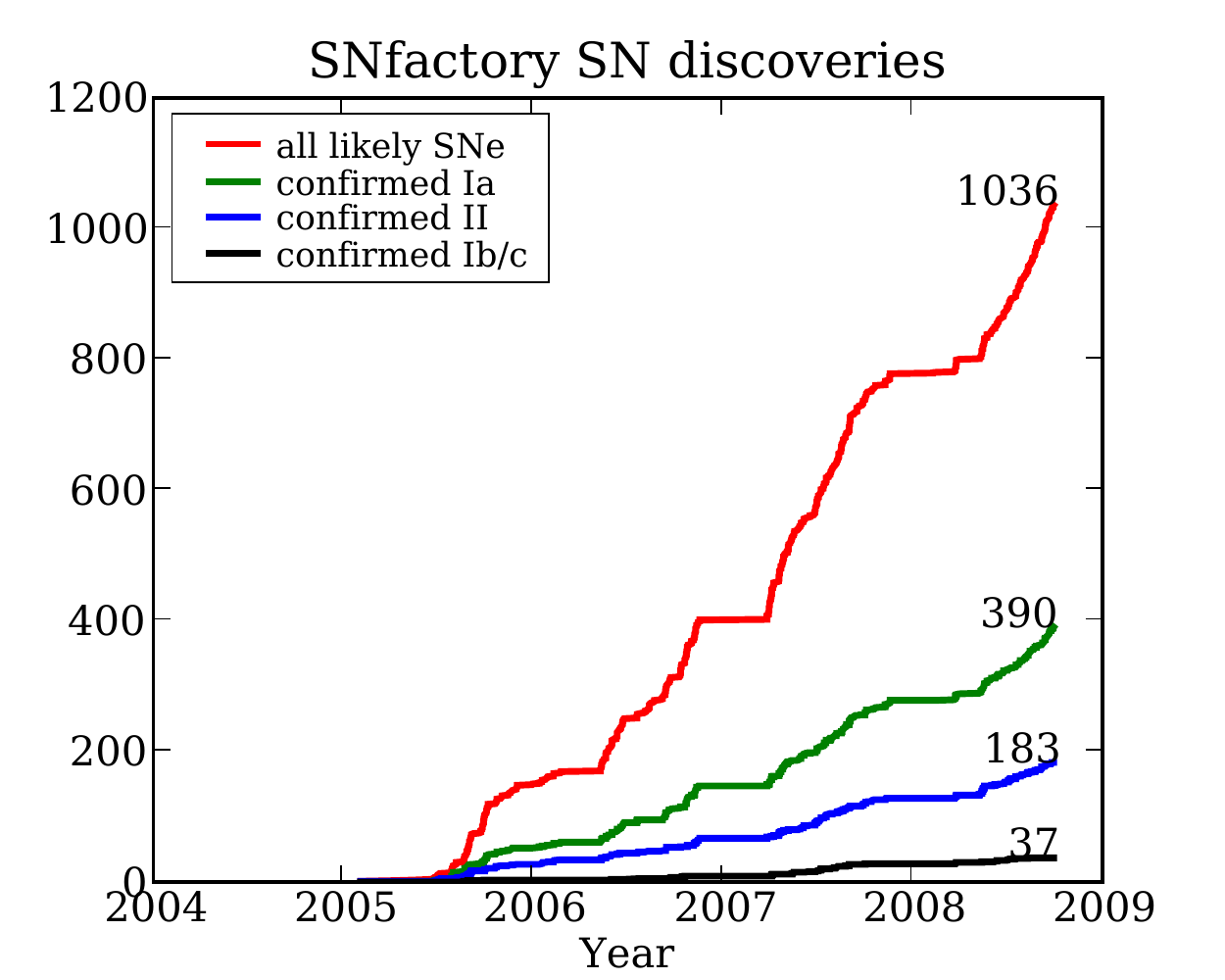}
\includegraphics[width=65mm]{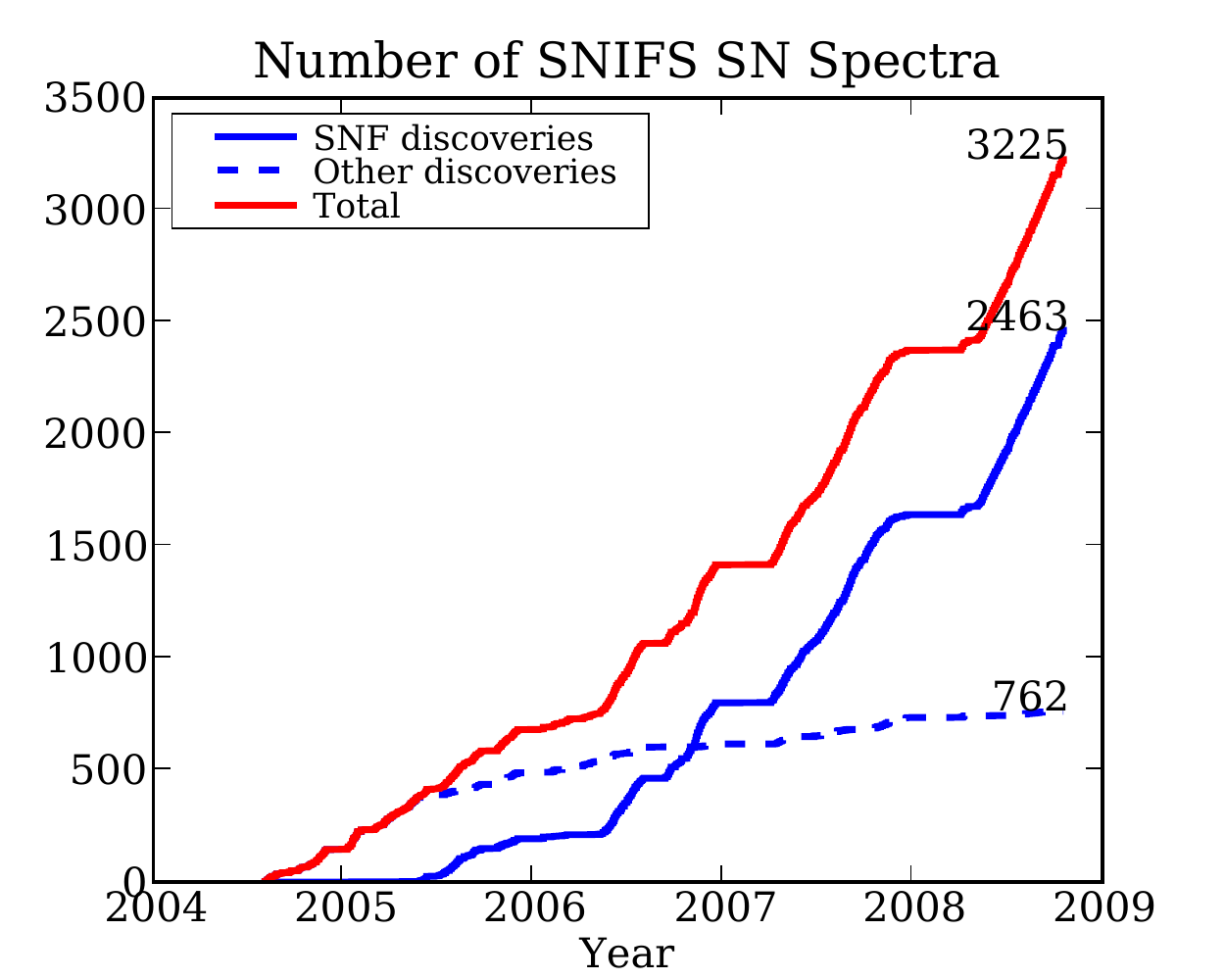}
\caption{Number of supernovae discovered (left) and 
supernova spectra obtained (right) by the Nearby Supernova Factory
as a function of time.  These plots include data obtained after
ICHEP 2008.  Flat regions reflect the Winter shutdowns each year
when we did not search or follow supernovae.} \label{fig:nSNeSpec}
\end{figure*}

Figure \ref{fig:nSNeSpec} (left) shows the number of supernovae
discovered as a function of time.  The SNfactory search finished
in September 2008 with over 1000 supernovae discovered, including
390 spectrally confirmed Type~Ia supernovae, 183 Type~II, and 37 Type~Ib/c.
The remainder have photometric confirmation as likely supernovae
but do not have a spectral type confirmation.
Figure \ref{fig:nSNeSpec} (right) shows the number of supernova spectra obtained
with SNIFS as a function of time.  Approximately 3/4 of these spectra are
from 185 supernovae for which we have a spectral timeseries.

The Nearby Supernova Factory search is the largest supernova search
ever performed in terms of both sky area and data volume.  The computational
scaling and false-positive object identification issues faced by the
SNfactory are relevant to the upcoming transient search pipelines of
the Palomar Transient Factory, 
PanSTARRS, 
and LSST, 
who intend to generate transient alerts within minutes of first
discovery \cite{baileyclassifiers}.




The unique screening and followup instrument SNIFS was
custom designed by the SNfactory
for the purpose of observing nearby supernovae.
Members of the Nearby Supernova Factory collaboration remotely
operate this telescope every 2 to 3 nights \cite{PierreSPIE}.
Within SNIFS, a photometric channel is used for initial target field acquisition,
telescope guiding during an exposure, and for monitoring the field stars
around a target to calibrate the nightly atmospheric extinction.
A prism redirects the light from a 6x6 square arcsecond field around the
supernova,
followed by a dichroic which splits this light to separate
blue (320 -- 520 nm) and red (510 -- 1000 nm) spectrograph channels.
Within each channel, a 15 x 15 lenslet array focuses the light which is
then dispersed into individual spectra.  These 225 spectra form a
datacube of spectra {\it vs.} position of
the supernova and its surrounding field.  These individual spectra are
used to deconvolve the contributions from the supernova, its host galaxy,
and the night sky background.
The lenslets capture all the light from the supernova,
enabling absolute flux calibration of the spectra when
combined with observations of spectrophotometric standard stars
and the surrounding field stars from the photometric channel.

This absolute flux calibration of spectral timeseries
is a unique feature of these supernova observations.
It allows synthesized photometry in any filter from 350 nm to 950 nm,
eliminating the significant systematic which arises from intercalibrating
the different filtersets used by current nearby and distant supernova
observations.  These data also contain far more information than the
datasets used to build current SN lightcurve models; it will enable
more accurate models, reducing another significant systematic.

\section{SNFACTORY DATA ANALYSES}


\begin{figure*}[t]
\centering
\includegraphics[width=80mm]{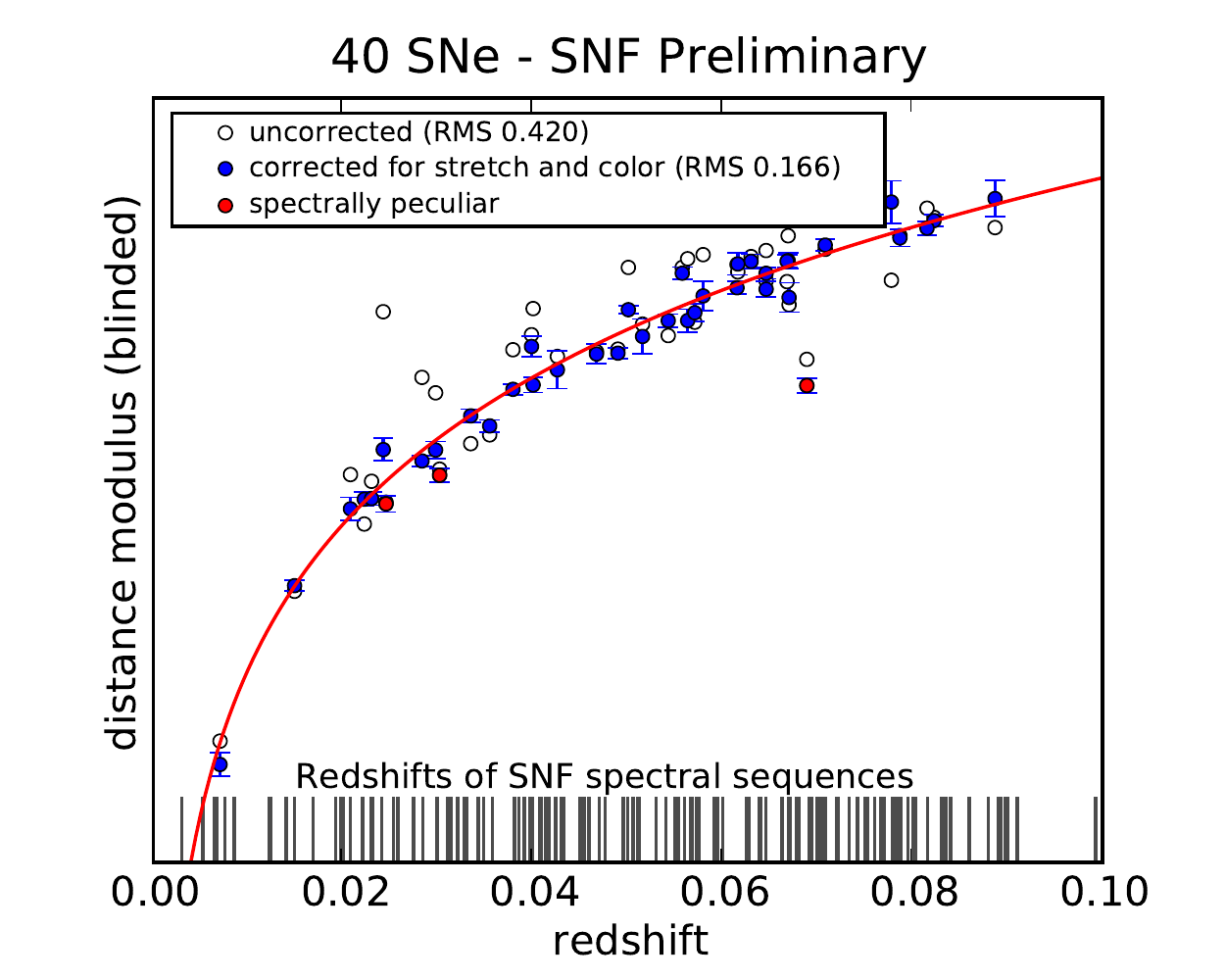}
\caption{Preliminary SNfactory nearby Hubble diagram with 40 supernovae.
Open circles show the uncorrected peak magnitudes; filled circles show
peak magnitudes after corrections for color and lightcurve stretch.
Spectrally peculiar supernovae are highlighted in red.  The redshifts
of additional SNfactory spectral timeseries measurements are also marked.
} \label{fig:hubble}
\end{figure*}

Figure \ref{fig:hubble} shows a Hubble diagram with 40 supernovae from the
Nearby Supernova Factory.  This is $\sim1/4$ of the total SNfactory dataset
of supernovae with spectral timeseries.  The remaining supernovae await
additional data processing, final reference observations, and/or
improved algorithms under development for disentangling host galaxy
backgrounds with complex spatial structure.
Before unblinding the cosmology we are performing additional systematics
cross checks, improving the data extraction 
and calibration algorithms, and quantifying the full error chain including
remaining systematics.
Before the absolute flux calibration is finalized, a variety of analyses
may be performed using the relative calibration (and thus the relative
residuals of the nearby Hubble diagram).  A few of these are highlighted
below.

Three spectrally unusual SNe in figure \ref{fig:hubble} are highlighted in red.
The largest outlier is SNF20070825-001, a SN whose spectra are a close
match to the super-Chandrasekhar mass supernova SNLS-03D3bb \cite{snls-superc}.
The SNfactory observations of this
target are the first spectral timeseries measurements of this class of
supernova; we have also discovered and observed another supernova
with similar spectra.  The two other highlighted SNe have spectra similar to
SN1991T, a known subclass of SNe~Ia which are brighter than typical.
These 3 supernova are well fit by the SALT2 lightcurve model yet their
absolute magnitudes are not well corrected by standard
stretch\footnote{In \cite{p99},
``stretch'' has a specific technical meaning relating to the Philips
relationship \cite{philips} that broader lightcurves tend to have brighter
peak magnitudes.  Other SN fitters have different parameters which describe
the same effect ($\Delta M_{15}$ for MLCS2k2; $x_1$ for SALT2).  Here, ``stretch''
refers generically to this correction, not a specific definition or fitter.}
and color corrections.  We are studying their spectra to better understand
SNe~Ia intrinsic diversity and discover better ways of calibrating their
peak magnitudes.

There are several ratios of spectral features which are known to
correlate with supernova lightcurve width \cite{RSi, RSiS, snapshotdistances}.
They may enable improved brightness calibration, but
their correlation with absolute magnitude has only been studied with
small statistics and it has never been
established whether they contain information beyond what is contained
in lightcurve shape and color.
Using the SNfactory dataset, large statistics correlations of these metrics
with absolute magnitude can be directly measured.
Most importantly, this dataset enables for the
first time a direct correlation of spectral indicators
with Hubble diagram residuals, {\it i.e.}, testing correlations
for information beyond stretch and color.
A preliminary analysis was done as a Master's thesis \cite{nico}; this
work is now being expanded with larger statistics and improved Hubble
diagram residual measurement errors.

Classic spectral metrics focus on specific spectral features and were usually
discovered by their correlation with stretch (or equivalent) rather than
what would be most cosmologically useful --- a correlation which contains
information beyond that of stretch and color.  The SNfactory is performing
a generalized correlation analysis of our supernova spectra, focusing on
features which correlate with absolute magnitude in ways which stretch
and color do not.  The results of this study will be reported at an
upcoming conference.


Previous SNfactory publications have contributed to the understanding
of SNe~Ia progenitor environments and the physics of their explosions
\cite{05gj, 06D}.
Ongoing studies include observations of the SNfactory discovered SN
SNF20080720-001 \cite{dusty},
the most reddened normal SN~Ia ever observed.
An analysis of this supernova is being performed using a spectroscopic ``twin''
with less reddening to disentangle the contributions from dust {\it vs.}
intrinsic SN properties.
We have also undertaken a program to obtain measurements of host galaxy
properties to study the connection between
SN~Ia properties and their stellar environments.

\begin{acknowledgments}

We are grateful to the technical and scientific staff of the University
of Hawaii 2.2-meter telescope for their assistance in obtaining these
data.  The authors wish to recognize and acknowledge the very
significant cultural role and reverence that the summit of Mauna Kea has
always had within the indigenous Hawaiian community.  We are most
fortunate to have the opportunity to conduct observations from this
mountain.
This work was supported in part by the Director,
Office of Science, Office of High Energy and Nuclear Physics, of the
U.S. Department of Energy under Contract No. DE-FG02-92ER40704, by a
grant from the Gordon \& Betty Moore Foundation, by  National Science
Foundation Grant Number AST-0407297, and in France by support from
CNRS/IN2P3, CNRS/INSU and PNC.  This research used resources of the
National Energy Research Scientific Computing Center, which is supported
by the Office of Science of the U.S.  Department of Energy under
Contract No. DE-AC02-05CH11231.  We also acknowledge support from the
U.S. Department of Energy Scientific Discovery through Advanced
Computing program under Contract No. DE-FG02-06ER06-04.
The search data was transferred using the High Performance Wireless
Research and Education Network (HPWREN), 
funded by the National Science Foundation grants 0087344 and 0426879.
SNIFS data were processed using resources from the CCIN2P3 computer
center supported by the IN2P3/CNRS.

\end{acknowledgments}

\newcommand\apj{ApJ}      
\newcommand\apjl{ApJ}     
\newcommand\pasp{PASP}
\newcommand\aap{A\&A}     
\newcommand\aj{AJ}        
\newcommand\nat{Nature}   
\newcommand\procspie{Proc.~SPIE}



\begin{thebibliography}{99} 


\bibitem{scpunion}
Kowalski, M., et al.\ 2008, 2008arXiv0804.4142K, accepted by ApJ

\bibitem{questII}
Baltay, C., et al.\ 2007, \pasp, 119, 1278 


\bibitem{snifs}
Aldering, G., et al.\ 2002, \procspie, 4836, 61 

\bibitem{baileyclassifiers}
Bailey, S., Aragon, C., Romano, R., Thomas, R.~C., Weaver, B.~A., 
\& Wong, D.\ 2007, \apj, 665, 1246 

\bibitem{PierreSPIE}
Antilogus, P., et al.\ 2008, \procspie, 7016,  





\bibitem{snls-superc}
Howell, D.~A., et al.\ 2006, \nat, 443, 308 

\bibitem{p99}
Perlmutter, S., et al.\ 1999, \apj, 517, 565 

\bibitem{philips}
Phillips, M.~M.\ 1993, \apjl, 413, L105 

\bibitem{RSi}
Nugent, P., Phillips, M., Baron, E., Branch, D., \& Hauschildt, P.\ 1995,\
\apjl, 455, L147 

\bibitem{RSiS}
Bongard, S., Baron, E., 
Smadja, G., Branch, D., \& Hauschildt, P.~H.\ 2006, \apj, 647, 513 

\bibitem{snapshotdistances}
Riess, A.~G., Nugent, P., 
Filippenko, A.~V., Kirshner, R.~P., \& Perlmutter, S.\ 1998, \apj, 504, 935 

\bibitem{nico}
Chotard, N. 2008,
``Etude d'indicateurs spectraux sur les spectres 
 de SN Ia issus de la collaboration SNFactory'',
Master's Thesis, IPNL Lyon, France


\bibitem{05gj}
Aldering, G., et al.\ 2006, \apj, 650, 510 

\bibitem{06D}
Thomas, R.~C., et al.\ 2007, \apjl, 654, L53 


\bibitem{dusty}
Aldering, G., et al.\ 2008, The Astronomer's Telegram, 1624, 1 

\end{thebibliography}
\end{document}